\documentclass[12pt]{article}
\usepackage{float}
\usepackage{epsfig}
\usepackage{amsmath}
\usepackage{url}
\usepackage{rotating}
\usepackage{setspace}
\usepackage[margin=0.85in]{geometry}
\usepackage{natbib}
\usepackage{authblk}
\usepackage{lineno}
\usepackage{footnote}
\usepackage{subcaption}
\usepackage{caption}
\usepackage{hyperref}
\hypersetup{draft}

\title{Defmod - Parallel multiphysics finite element code for modeling crustal deformation during the earthquake/rifting cycle}
\author{S. Tabrez Ali}
\affil{Department of Geoscience\\University of Wisconsin\\Madison, WI 53706, USA}
\date{December 25, 2015}
\begin{document}
\maketitle


\begin{abstract}
In this article, we present Defmod, an open source, fully unstructured, two or three dimensional, parallel finite element code for modeling crustal deformation over time scales ranging from milliseconds to thousands of years. Unlike existing public domain numerical codes, Defmod can simulate deformation due to all major processes that make up the earthquake/rifting cycle, in non-homogeneous media. Specifically, it can be used to model deformation due to dynamic and quasistatic processes such as co-seismic slip or dike intrusion(s), poroelastic rebound due to fluid flow and post-seismic or post-rifting viscoelastic relaxation. It can also be used to model deformation due to processes such as post-glacial rebound, hydrological (un)loading, injection and/or withdrawal of fluids from subsurface reservoirs etc. Defmod is written in Fortran 95 and uses PETSc's parallel sparse data structures and implicit solvers. Problems can be solved using (stabilized) linear triangular, quadrilateral, tetrahedral or hexahedral elements on shared or distributed memory machines with hundreds or even thousands of processor cores. In the current version of the code, prescribed loading is supported. Results are written in ASCII VTK format for easy visualization. The source code is released under the terms of GNU General Public License (v3.0) and is freely available from \url{https://bitbucket.org/stali/defmod/}. 
\end{abstract}

\section{Introduction}

Computer codes are key for simulating crustal deformation due to earthquakes, volcanic intrusions, hydrological loading and anthropogenic activity. By simulating deformation, in combination with geodetic and seismic observations, we can gain insight into underlying geophysical processes and estimate parameters of interest. Unstructured mesh based codes, such as those based on the finite element method, allow for flexibility in distribution of material properties and geometry that is difficult to accommodate in analytical or semi-analytical codes. In this short article, we present Defmod, an open source, fully unstructured, parallel multiphysics finite element code that is build on top of PETSc \citep[Portable, Extensible Toolkit for Scientific computation,][]{petsc}, a modern, scalable numerical library for sparse linear algebra that provides a suite of parallel sparse data structures, linear solvers and preconditioners. Defmod is written in just $\sim2000$ lines of Fortran 95 and is easy to adapt and extend, thus making it useful not only for research but also for learning and teaching. Unlike existing public domain numerical codes, which can simulate deformation due to some of the processes that make up earthquake/rifting cycle, Defmod can model all the major processes in non-homogeneous media, i.e., co-seismic slip or dike intrusion(s), poroelastic rebound and viscoelastic relaxation, using (stabilized) linear finite elements.




\section{Physics and Implementation}
The fundamental equation that Defmod solves is the Cauchy's equation of motion:
\begin{equation}
\int_{\Omega} (\sigma_{ij,j}+f_{i}-\rho \ddot{u}_{i})d\Omega = 0
\end{equation}
where $\sigma$ is the stress tensor, $f$, the body force, $\rho$, the density, $u$, the displacement field and $\Omega$ the control volume. Using the finite element method, the equation can be written in the semi discrete form as:
\begin{equation}
M\ddot{u}+C\dot{u}+Ku=f
\end{equation}
\citep{ocz2k}, where $M$ is the mass matrix, $C$, the damping matrix and $K$, the stiffness matrix. To impose fault slip/opening and displacement/velocity boundary conditions, Defmod uses linear constraint equations. The constraints are implemented using Lagrange Multipliers, which results in the system of equations:
\begin{equation}
M\ddot{u}+C\dot{u}+Ku+G^{T}\lambda=f
\label{semid}
\end{equation}
\begin{equation}
Gu=l
\label{lm}
\end{equation}
where $G$ is the constraint matrix, $\lambda$ is the force required to enforce the constraint and $l$ is the value of the constraint (e.g., prescribed slip or displacements).

For dynamic wave propagation problems, Defmod uses an explicit solver that performs time integration using a central difference scheme for acceleration $\ddot{u}$, and a backward difference scheme for velocity $\dot{u}$. The displacements at time step $t+\Delta t$ are calculated using the equation:
\begin{equation}
u^{t+\Delta t} = M^{-1}\left[\Delta t^{2}(f-Ku^{t})-\Delta tC(u^{t}-u^{t-\Delta t})\right]+2u^{t}-u^{t-\Delta t}
\end{equation}
The mass matrix $M$ is assumed to be lumped, therefore is diagonal, and easy to invert. A common choice for $C$ is ``proportional'' or Rayleigh damping in which:   
\begin{equation}
C=\alpha M+\beta K
\end{equation}
where $\alpha$ and $\beta$ are user supplied damping coefficients. Due to the Courant-Friedrichs-Lewy criteria, the critical time step $\Delta t_{critical}$ in the explicit scheme is restricted by the equation:
\begin{equation}
\Delta t_{critical} \leq L/c;~c=\sqrt{E/\rho}
\end{equation}
where $L$ is the length of the smallest element, $E$ is the Young's modulus and $\rho$ is the density of the element. For problems with constraints, Defmod uses the ``Forward Increment Lagrange Multiplier Method'' of \citet{carpenter}. To absorb waves at boundaries, the local element level scheme proposed by \citet{lysmer} is used. It consists of a series of dashpots placed normally and tangentially at the boundary nodes. This completely absorbs waves that approach the boundary at a normal incidence angle. For oblique angles of incidence, or for evanescent waves, the energy absorption is not perfect.

In quasistatic viscoelastic problems, the inertial terms in equations \ref{semid} and \ref{lm} are neglected and the resulting implicit, indefinite system of equations can be solved using a parallel, sparse direct or preconditioned iterative solver, specified at run-time. The implicit time stepping algorithm for viscoelastic relaxation is based after \citet{tekton}. 



\begin{figure*}[t]
\centering
\includegraphics[width=5in]{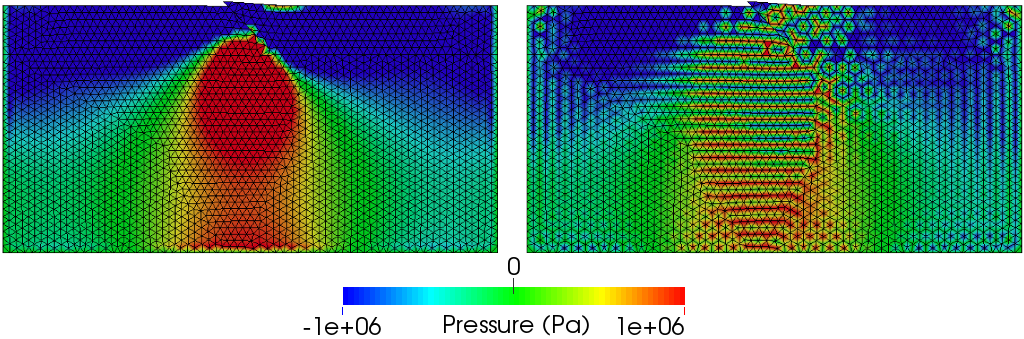}
\caption{Pore pressure field inside a two dimensional poroelastic domain, discretized using linear triangular elements, following co-seismic slip on a thrust fault, with (left) and without (right) stabilization.}
\label{stable}
\end{figure*}

In quasistatic poroelastic problems, we also have to solve the continuity equation, in addition to the momentum equation. This results in a coupled system of equations, of the form:
\begin{equation}
K_{e}u-Hp=f
\end{equation}
\begin{equation}
H^{T}\dot{u}+S\dot{p}+K_{c}p=q
\end{equation}
\citep{ocz2k}, where $K_{e}$ and $K_{c}$ are solid and fluid stiffness matrices, $H$ is the coupling matrix, $S$ the compressibility matrix, $p$ the pressure vector and $q$ the in/out flow. The time-dependent system of equations is solved using an incremental loading scheme \citep{smith}. To circumvent the Ladyzenskaja-Babuska-Brezzi restrictions on linear elements, Defmod uses the local pressure projection scheme proposed by \cite{dohrman}. The scheme works well for linear quadrilateral and hexahedral elements \citep{white} but its use for linear triangles and tetrahedrons has not been demonstrated in the peer reviewed literature for poroelasticity problems. The stabilization works well as long as a higher order integration scheme is used, i.e., 3-point integration for triangles and 4-point integration for tetrahedral elements. For example, Figure \ref{stable} shows the pressure field inside a two dimensional poroelastic domain, discretized using linear triangular elements, following co-seismic slip on a thrust fault, with and without stabilization. By using stabilized linear elements, Defmod can solve poroelasticity problems more efficiently than codes that use quadratic approximation for displacements and linear for pressure. 

\begin{figure*}[ht]
\centering
\includegraphics[width=3.5in]{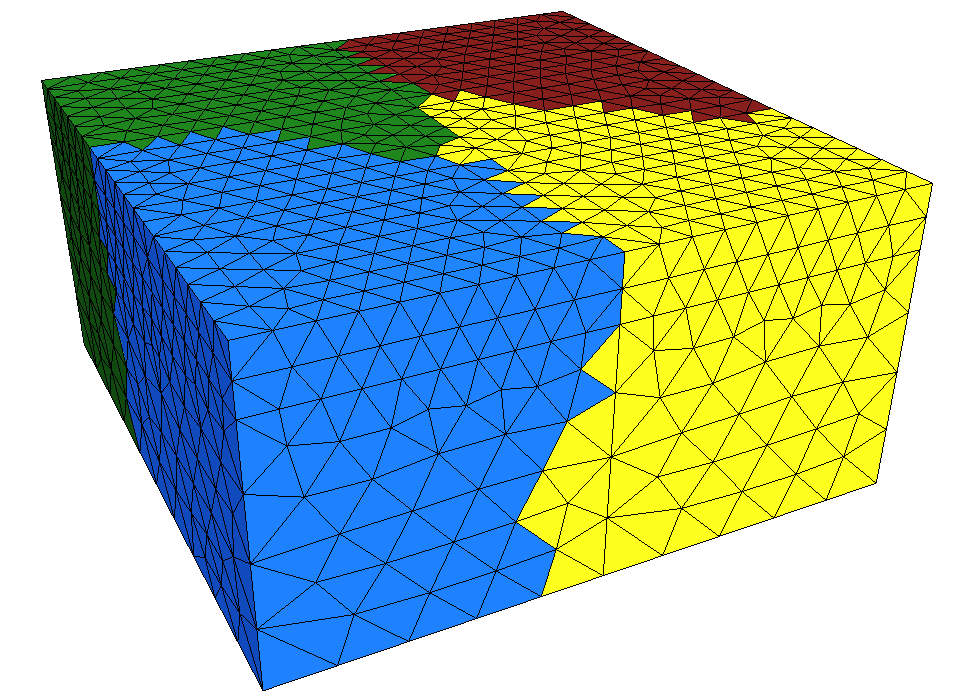}
\caption{A three dimensional mesh, made of linear tetrahedral elements, partitioned across 4 processor cores.}
\label{partition}
\end{figure*}
\begin{figure*}[h]
\centering
\includegraphics[width=3.5in]{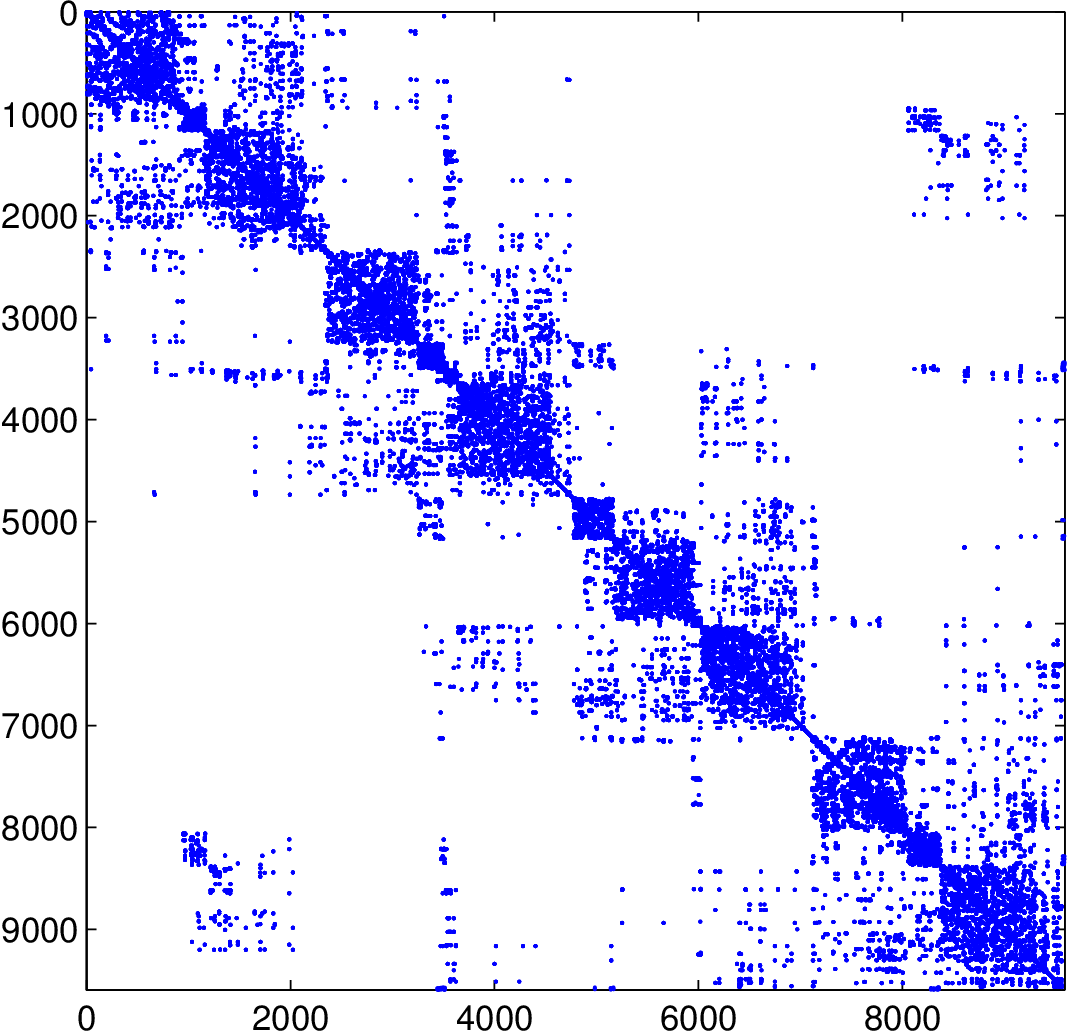}
\caption{Stiffness matrix $K$ resulting from the mesh, shown in Figure \ref{partition}, distributed across 4 processor cores. Each core owns a fixed number of rows ($\sim$2400 for the example shown above) of the matrix as well as the rows of corresponding vectors of the linear system. Each blue dot corresponds to a non-zero entry in the sparse matrix. Fewer entries in the off-diagonal block(s) help to minimize MPI communication during assembly and linear algebra operations such as sparse matrix-vector multiplication.}
\label{matviz}
\end{figure*}

Parallelism in Defmod is achieved via domain decomposition, i.e., each rank (or processor core) owns a subset of elements and nodes, and a subset of the global sparse matrices and vectors, partitioned row wise by PETSc. All element level calculations, e.g., formation of local matrices (e.g., $K$ and $M$), recovery of stress etc., therefore are local to the rank. Assembly and linear algebra operations on distributed matrices and vectors such as sparse matrix-vector multiplication, however, requires MPI\footnote{via Message Passing Interface \citep{mpi}} communication between ranks which is minimized using efficient mesh partitioning and renumbering of nodes. Figure \ref{partition} shows a mesh that has been partitioned across 4 processor cores using METIS \citep{metis}. The corresponding stiffness matrix $K$ is shown in Figure \ref{matviz}. MPI communication is also required to update the solution at ``ghost'' nodes at each time step. Because all data on a processor core prior to the ``solve'' phase, is local, Defmod can easily solve large problems, i.e., with hundreds of millions of unknowns. 

\section{Availability, Installation and Usage}
\begin{figure*}[h]
\centering
\includegraphics[width=4.25in]{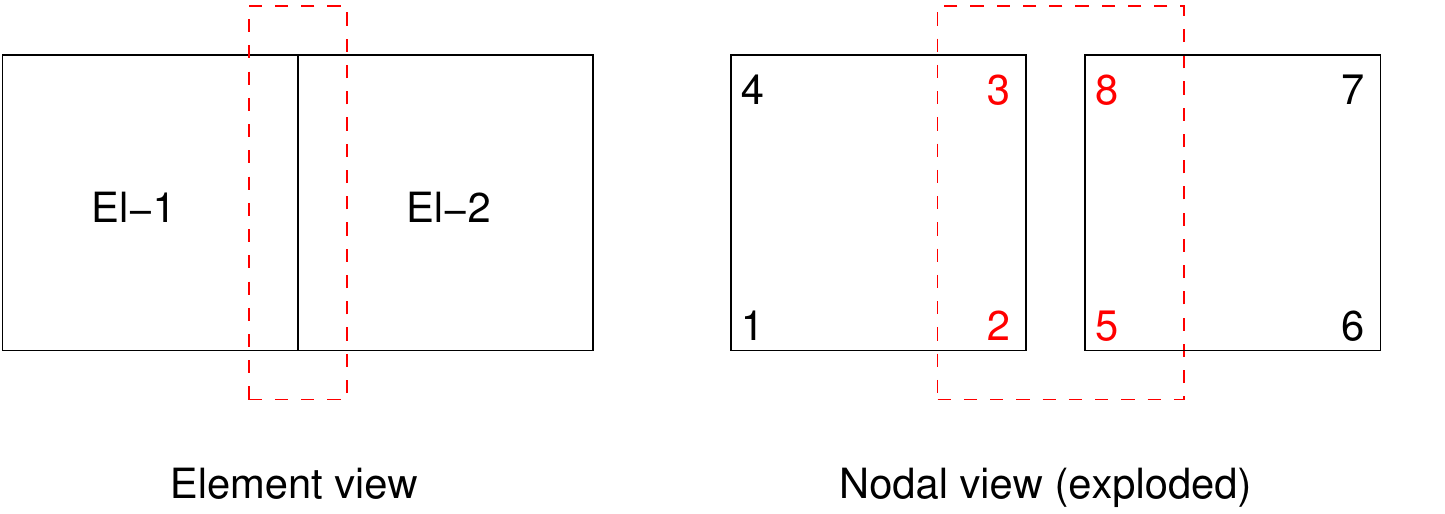}
\caption{Exploded view of the fault interface showing coincident node pairs ($\{3,8\}$ and $\{2,5\}$) that accommodate slip/opening.}
\label{fault}
\end{figure*}

To compile Defmod, PETSc (freely available from \url{http://www.mcs.anl.gov/petsc/}), must first be configured and installed along with METIS. Instructions for compiling PETSc, and subsequently Defmod, are included with the source. On some Linux distributions, e.g., Debian, prebuilt PETSc binaries are available from the distribution repository. Once PETSc is installed and the \texttt{PETSC\_DIR} variable set, Defmod can be downloaded and compiled simply by executing the following commands:

\begin{verbatim}
   # hg clone https://bitbucket.org/stali/defmod
   # cd defmod
   # make all
\end{verbatim}

The code has been tested with a number of C/Fortran 95 compilers including those from Cray, GNU, Intel, Open64, PGI and Solaris Studio. Defmod uses a single ASCII input file that contains all simulation parameters, mesh information, i.e., nodal coordinates and element connectivity data, material data, time-dependent linear constraint equations, as well as any time-dependent force/flow and/or traction/flux boundary conditions. The input file also contains information about fixed/roller boundaries, Winkler foundation(s) for simulating effects of isostasy, as well as absorbing boundaries. To generate the mesh, i.e., nodal coordinates and connectivity data, external meshing packages such as Cubit \citep{cubit}, Gmsh \citep{gmsh} etc., must be used. A number of sample input files, for different types of simulations, are available with the source code, in the \texttt{defmod/examples} directory. Some of the files are commented and help new users to understand the file structure. To run a simulation, the \texttt{mpiexec} command or its equivalent must be used. For example:
\begin{verbatim}
   # mpiexec -n 2 ./defmod -f examples/two_quads_qs.inp
\end{verbatim}

Slip on faults can be modeled using coincident nodes, i.e., two nodes that share the same coordinate in space but belong to different elements, on either side of the fault. To enforce slip between two coincident nodes, linear constraint equations must be used. For example, to specify opening of 5 m between nodes 3 and 8, which belong to elements 1 and 2, respectively, as shown in Figure \ref{fault}, we use two equations: \texttt{Ux$_{3}$-Ux$_{8}$=5.0} and \texttt{Uy$_{3}$-Uy$_{8}$=0.0}, which can be specified in the input file, along with the time duration, during which the constraint is active. Constraint equations can also be used to make a fault permeable, e.g., equation \texttt{P$_{3}$-P$_{8}$=0.0} will ensure continuity of pressure across coincident nodes 3 and 8. Similarly, equations can be used for imposing time-dependent nonzero displacement and/or pressure boundary conditions for different kinds of loading scenarios with or without faults. 

Because Defmod uses PETSc, all of PETSc's command line options are also available to Defmod users. The list of options can be obtained by using the \texttt{-help} switch. By default, quasistatic problems are solved using ASM (Additive Schwarz Method) preconditioned GMRES (Generalized Minimal RESidual) method. For problems without constraints, the Conjugate Gradient (CG) method can be used instead, by specifying the command line option $\texttt{-ksp\_type cg}$. If PETSc is configured and installed with a parallel sparse direct solver such as MUMPS \citep[MUltifrontal Massively Parallel Solver,][]{mumps}, then it can be engaged simply by using the command line option $\texttt{-pc\_type~lu}$ $\texttt{-pc\_factor\_mat\_solver\_package~mumps}$. Users are encouraged to experiment with various solvers available through PETSc. By default, each rank writes its own output in ASCII VTK format that can be directly visualized using packages such as ParaView \citep{paraview}, VisIt \citep{visit} etc. Users can also easily manipulate the output using standard Unix/Linux shell utilities. Figure S\ref{viz} shows some results obtained using Defmod.

\section{Validation}

To validate Defmod, we compare results obtained from Defmod to those obtained by Abaqus \citep{abaqus}, a commercial finite element code, for the equivalent elastodynamic and quasistatic problems. We use the same finite element meshes, generated using Cubit, in both codes\footnote{For the quasistatic poroelastic problem, quadratic elements were used in Abaqus where as linear elements were used in Defmod.}. This mitigates the effect of mesh resolution, time step, boundary/interface and loading conditions on the solution. In all cases we assume a Young's modulus of 30.0 GPa, and a Poisson's ratio of 0.25. 

\begin{figure*}[h]
\centering
\includegraphics[width=5.25in]{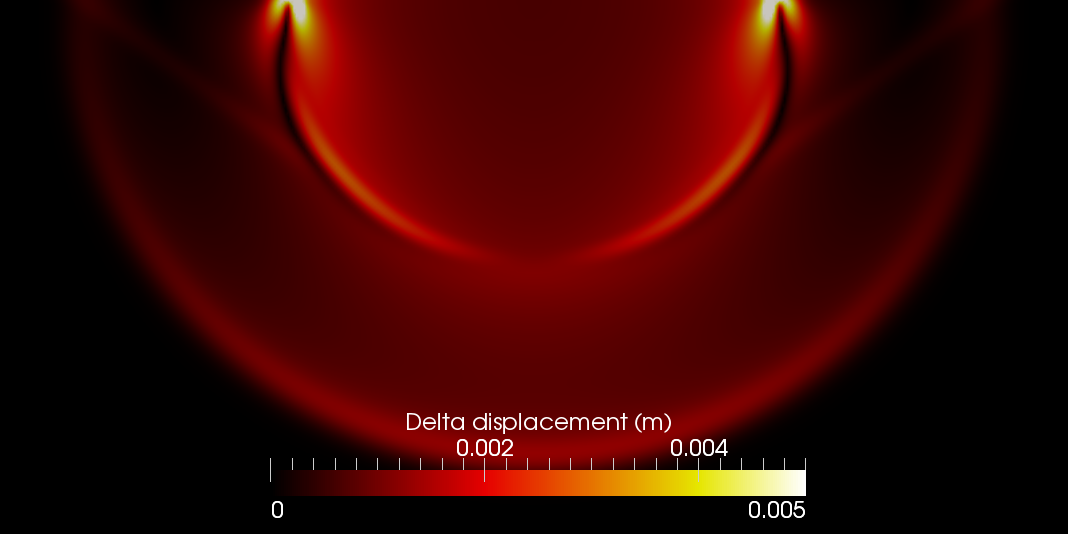}
\caption{Snapshot of the wave field, showing delta displacements ($\Delta u$), at the 250$^{th}$ time step ($\Delta t=0.05$ seconds).}
\label{case2fig}
\vskip 2.5mm
\hskip -10mm \includegraphics[width=3.75in]{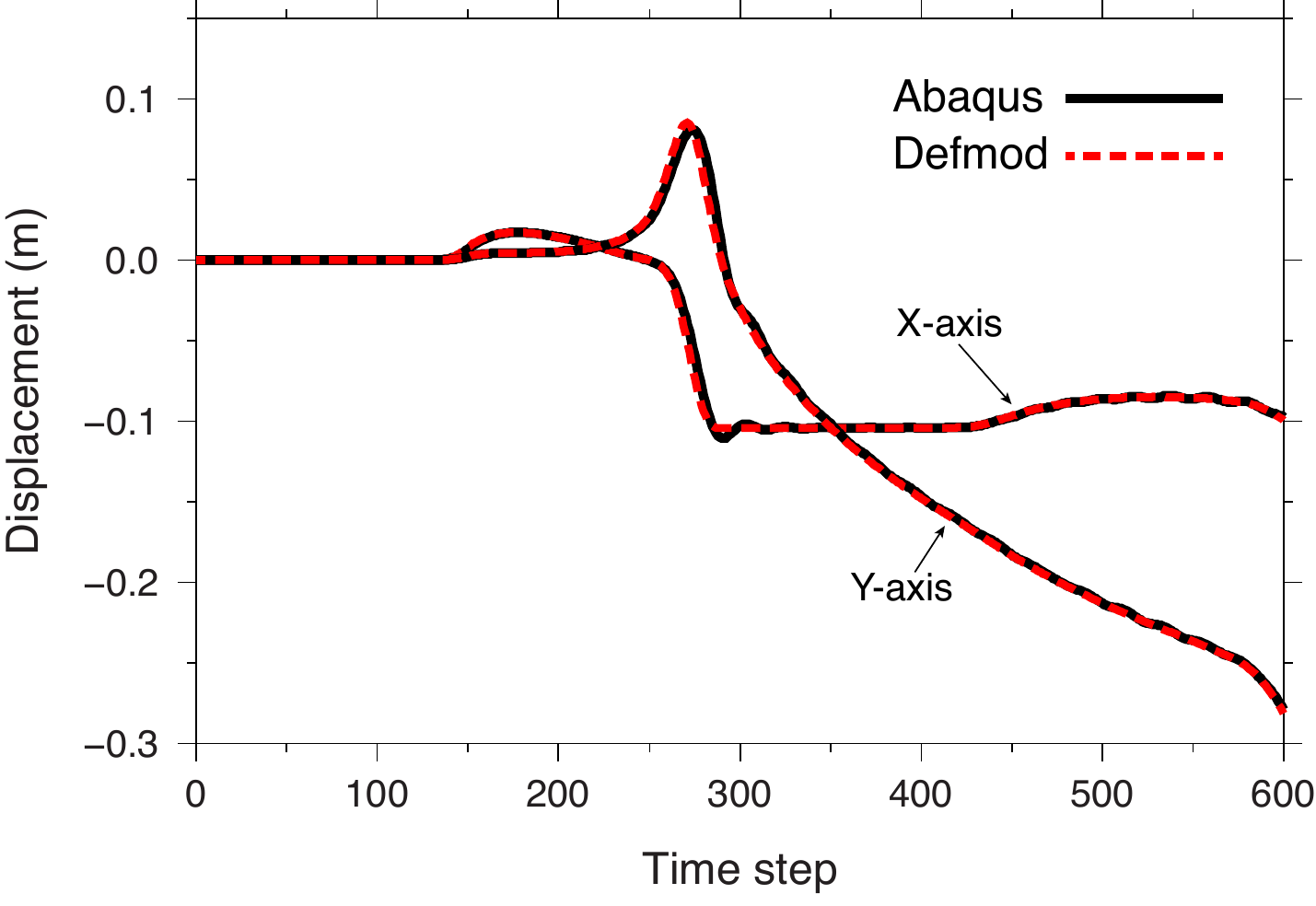}
\caption{Horizontal (X-axis) and vertical (Y-axis) displacements at a point on the surface, 25 km away from the impulsive source.}
\label{case2}
\end{figure*}

For the elastodynamic case we solve a variant of the Lamb's problem where we calculate the response of a 100 km by 50 km, two dimensional elastic domain that is discretized using linear quadrilateral elements, to an instantaneous point force (of -10.0 GN in the vertical direction) applied at the free surface. We calculate surface displacements at a distance of 25 km from the source. A stiffness proportional damping coefficient of $\beta=0.01$ is assumed. Figure \ref{case2fig} shows a snapshot of the wave field calculated using Defmod at the 250$^{th}$ time step. The vertical and horizontal displacements, calculated over time, using Defmod and Abaqus, are in good agreement with each other, as shown in Figure \ref{case2}. The small oscillations in the Abaqus solution are due to the use of a reduced integration scheme. Unlike Abaqus, Defmod uses full integration for the elastodynamic problem.

\begin{figure*}[h]
\centering
\hskip 15mm \includegraphics[width=3.0in]{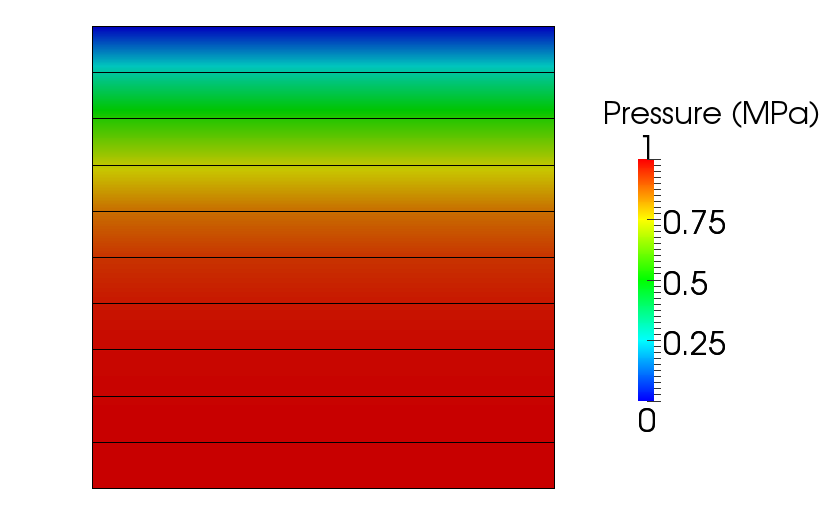}
\caption{Pore pressure field for the Terzaghi consolidation problem, at $t=1.0$ seconds.}
\label{case3fig}
\vskip 2.5mm
\hskip -10mm \includegraphics[width=3.75in]{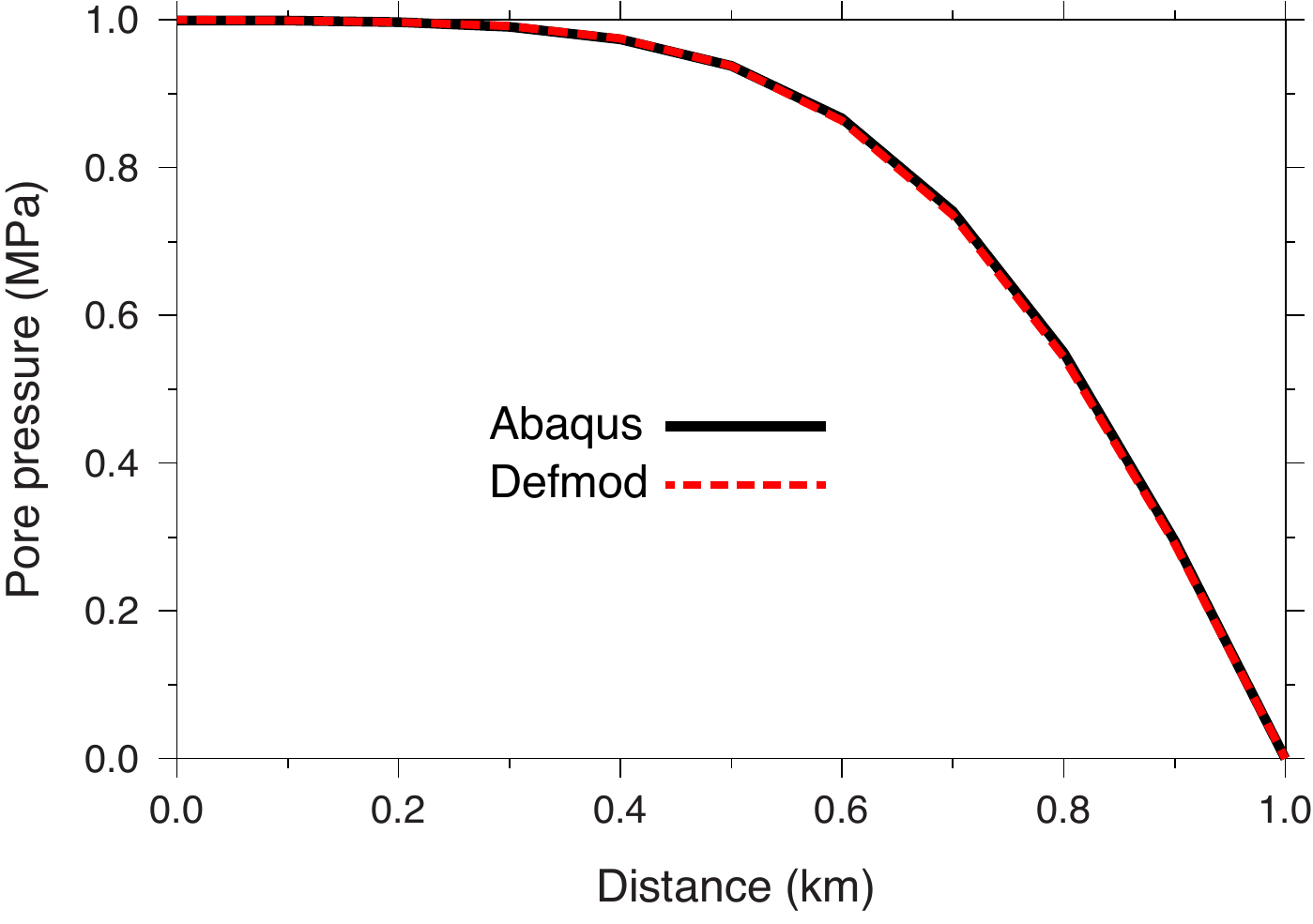}
\caption{Variation in pore pressure with depth, at $t=1.0$ seconds.}
\label{case3}
\end{figure*}

For the quasistatic poroelastic case, we solve the Terzaghi consolidation problem where we calculate the pore pressure change due to sudden application of distributed load. The model domain is 1 km long by 1 km wide. The nodes at the bottom boundary are fixed and vertical displacements are allowed on all other nodes. Fluid is allowed to flow through the top surface where a sudden load of 1.0 MPa is applied. A permeability of 1.0$^{-9}$ m$^{2}$ is assumed for the entire domain. The resulting pore pressure field, calculated using Defmod, at 1.0 seconds, is plotted in Figure \ref{case3fig}. Figure \ref{case3} shows the Defmod solution along with the Abaqus solution, along a vertical cross-section, at 1.0 seconds. Once again, we find excellent agreement between the two solutions. Unlike Abaqus, Defmod uses stabilized linear elements and therefore can solve for the pressure field more efficiently.

\begin{figure*}[h]
\centering
\includegraphics[width=5.5in, trim = 0mm 8mm 0mm 16mm,clip]{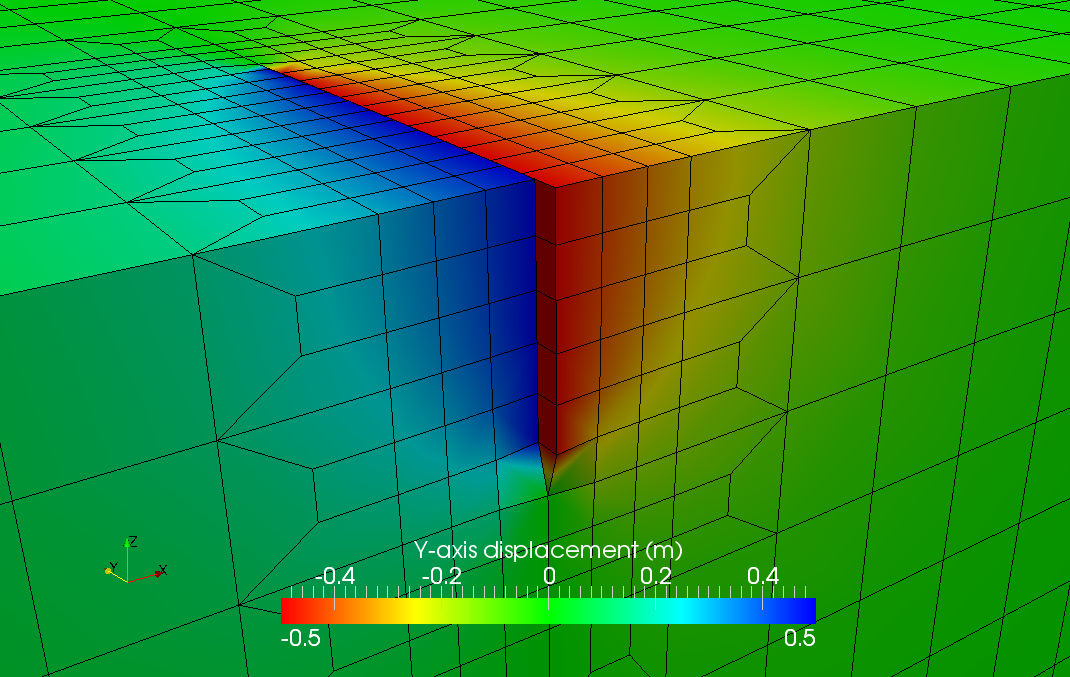}
\caption{Part of the finite element mesh used for validation of quasistatic viscoelastic relaxation. Colors represent fault parallel displacements, at $t=0$ years.}
\label{case1fig}
\end{figure*}

\begin{figure*}[h]
\centering
\hskip -10mm \includegraphics[width=3.75in]{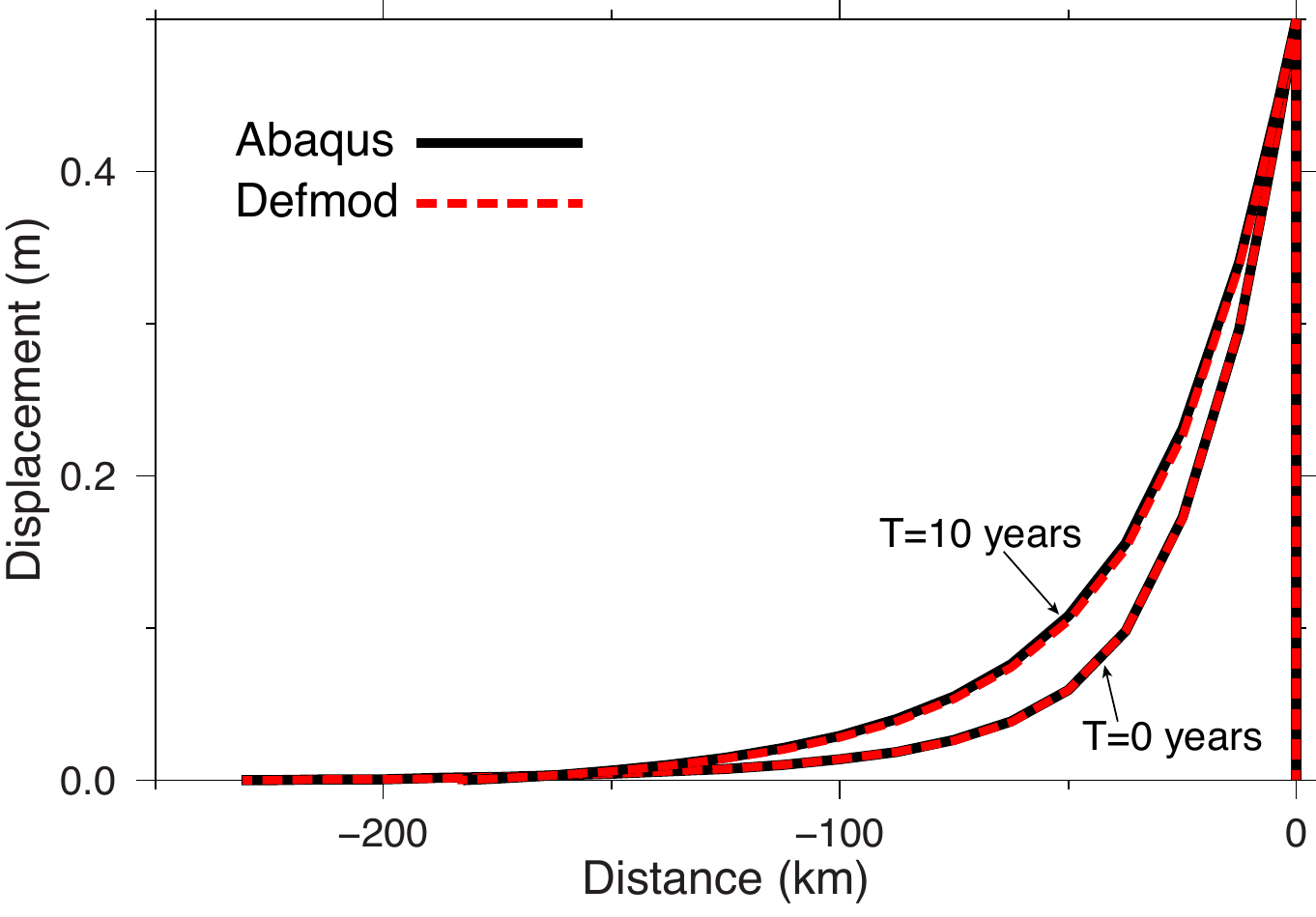}
\caption{Fault parallel surface displacements across the fault, in the -X direction.}
\label{case1}
\end{figure*}

For the quasistatic viscoelastic case, we model deformation due to 1.0 m of prescribed slip on a 100 km long strike-slip fault that extends from the surface to a depth of 25 km. The elastic crust is assumed to be 25 km thick and overlies a 225 km thick viscoelastic mantle that has a Maxwell viscosity of $\eta=10^{18}$ Pa-s. The model domain is 500 km by 500 km by 250 km and is discretized using three dimensional linear hexahedral elements, part of which is shown in Figure \ref{case1fig}. The Defmod solution at the surface, at $t=0$ and $t=10$ years, is plotted in Figure \ref{case1} along with the Abaqus solution, along a fault perpendicular cross-section. Once again, we find good agreement between the two solutions. Minor discrepancies could be due to differences in the underlying algorithms and their implementation within the two codes.

\section{Parallel Performance}
We demonstrate parallel performance of Defmod's implicit and explicit solvers using ``strong scaling'', i.e., the number of processor cores is increased for a problem of fixed size. To do so, we solve a three dimensional variant of the Lamb's problem that contains $\sim$25 million unknowns. The reason for choosing such a simple problem is because it includes calculations that are common to all types of problems, irrespective of rheology (i.e., elastic, poroelastic/viscoelastic or poroviscoelastic), rheology contrasts, force/flow boundary conditions, use of constraint equations (e.g., presence or absence of faults or non-zero displacement and/or pressure boundary conditions etc.) and so on. We solve the problem on 128, 256, 512 and 1024 processor cores of the Trestles cluster at the San Diego Supercomputer Center. Each node in the cluster has 4, 2.4 GHz 8-core AMD Magny-Cours processors and 64 GB of RAM. The nodes are interconnected in a fat-tree topology via a quad-data-rate InfiniBand fabric. For benchmarking the solver performance, we exclude the time taken for the initialization phase, i.e., reading and partitioning of the mesh. However, timings for all other calculations, including formation of local matrices and vectors, parallel assembly and solution of the linear system are included. Because the number of iterations in the iterative solver for quasistatic problems can slightly vary with the number of processor cores, we fix the number of iterations. Results for both, explicit elastodynamic and implicit quasistatic problems are shown in Table \ref{table}.

Overall, we find very good speedup, i.e., $\sim$0.90 or higher, for both problems. For the explicit elastodynamic problem, parallel efficiency decreases relatively fast (from 0.93 to 0.80) when moving from 512 to 1024 processor cores. This is because the amount of work done per processor core becomes too low. On most current generations clusters, we find that as long as degrees of freedom (DOF) per core is $\geq$ 37.5K, an efficiency of $\sim$0.90 can be achieved. The implicit solver for the quasistatic problem performs slightly better even with a lower DOF/core value, because of local element level calculations (i.e., recovery of stress, formation of the body force vector $f$ etc.) performed during each time step of the solve phase. We do not evaluate single node performance as the memory bandwidth available to each core on a socket depends on the number of cores being used per socket.

\begin{table*}[h]
\begin{center}
\begin{tabular}{  c | c | c | c  }\hline 
CPU cores & DOF per & Explicit elastodynamic  & Implicit quasistatic \\
used & core & solver efficiency & solver efficiency (residual)\\ \hline
128 & 195.2K & 1.00 & 1.00 (5.357e-05) \\ 
256 & 97.6K & 0.97 & 0.92 (5.212e-05) \\
512 & 48.4K & 0.93 & 0.89 (5.125e-05) \\
1024 & 24.4K & 0.80 & 0.87 (4.965e-05) \\\hline 
\end{tabular}
\end{center}
\caption{Solver performance for a problem consisting of 25 million degrees of freedom (DOF). The number in parenthesis represents the final preconditioned residual norm for the last time step at the 500th iteration.}
\label{table}
\end{table*}

\section{Conclusions}
In this short article, we present Defmod, a two or three dimensional, parallel multiphysics finite element code for modeling crustal over time scales ranging from milliseconds to thousands of years. It can be used to simulate deformation due to dynamic and quasistatic processes such as earthquakes, dike intrusions, poroelastic rebound, post-seismic or post-rifting viscoelastic relaxation, post-glacial rebound, hydrological (un)loading, injection and/or withdrawal of fluids from subsurface reservoirs etc. By using PETSc's parallel sparse data structures and implicit solvers, Defmod can solve problems, with hundreds of millions of unknowns, on large, shared or distributed memory machines with hundreds or even thousands of processor cores. By using stabilized linear elements, Defmod can solve poroelasticity problems more efficiently than codes that use higher order approximation for the displacement field. Because it is written in just $\sim2000$ lines of Fortran 95, it can easily be adapted and extended, making it useful not only for research but also for instructional purposes.


\section{Acknowledgements}
(i) PETSc team for providing a first class Fortran API and support, (ii) Gregory Lyzenga and Jay Parker for providing the three dimensional viscoelastic strain rate matrices, (iii) Christodoulos Kyriakopoulos for his help in validating Defmod, and (iv) Kurt Feigl and Herb Wang for support and constructive suggestions. Computing resources for benchmarking and testing Defmod were provided under XSEDE award TG-EAR110020. The author was partially supported by U.S. National Science Foundation award EAR0810134 and U.S. Department of Energy's Geothermal Technologies Office award \textnormal{DE-EE0005510} during the preparation of this article.

\bibliographystyle{apalike}
\bibliography{mybib}


\clearpage

\makeatletter
\renewcommand{\fnum@figure}{\figurename~S\thefigure}
\makeatother
\setcounter{figure}{0}

\begin{figure*}
\centering
\includegraphics[width=6.625in]{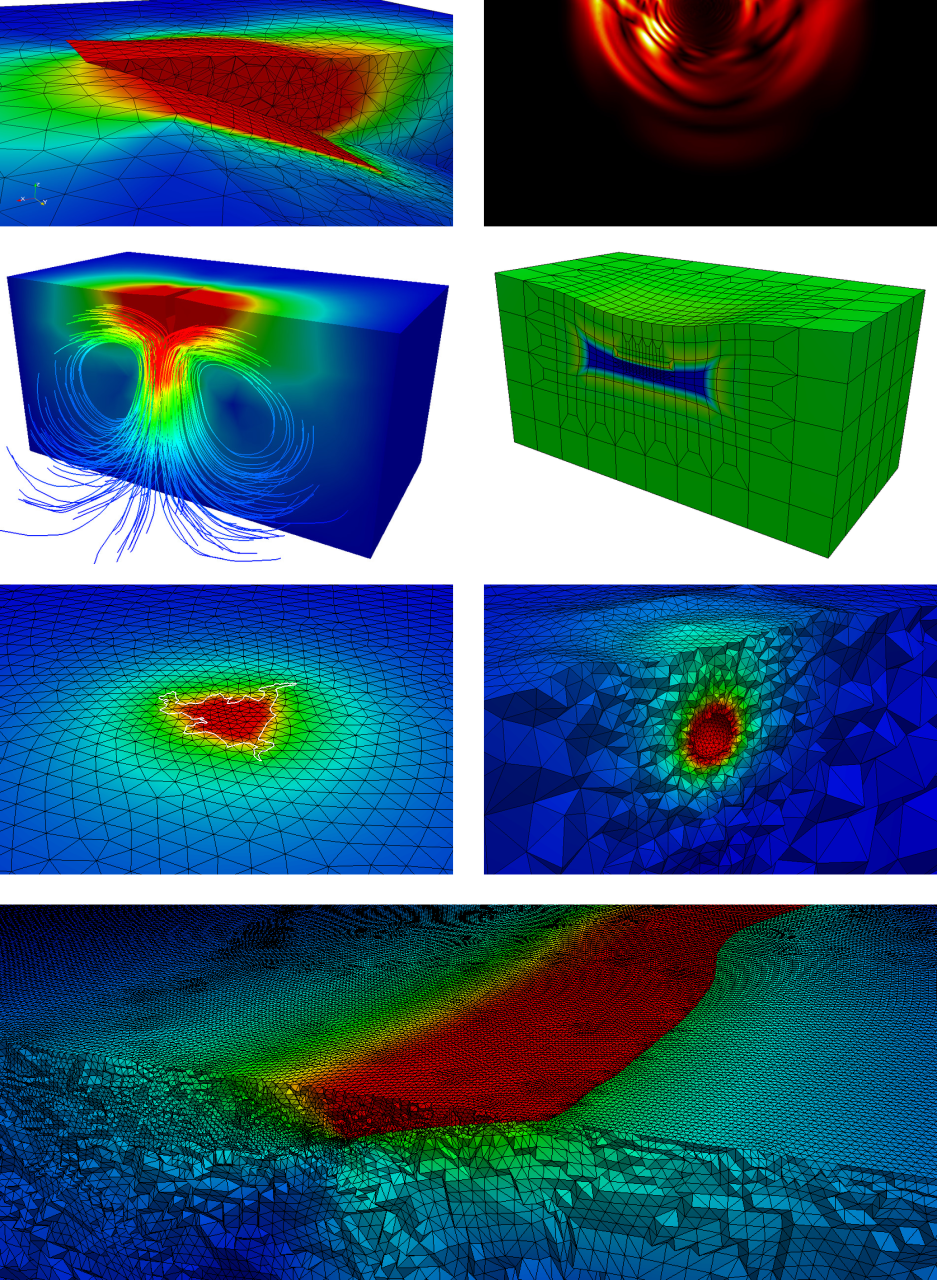}
\caption{Gallery of images generated using Defmod, visualized using ParaView.}
\label{viz}
\end{figure*}

\end{document}